\documentclass[conference]{IEEEtran}

\ifCLASSINFOpdf
   \usepackage[pdftex]{graphicx}
  \graphicspath{{../../EFTF/Torino2018/talk_PEP/Figures/}}
   \DeclareGraphicsExtensions{.pdf,.jpeg,.png}
\else
   \usepackage[dvips]{graphicx}
   \graphicspath{{../talk_PEP/Figures/}}
   \DeclareGraphicsExtensions{.eps}
\fi
\usepackage{amsmath}
\usepackage{url}
\usepackage[latin1]{inputenc}
\usepackage[T1]{fontenc}
\usepackage{color}
\usepackage{soul}

\begin{document}
\title{Two-branch fiber link for international clock networks}

\author{\IEEEauthorblockN{Dan Xu\IEEEauthorrefmark{1},
Etienne Cantin\IEEEauthorrefmark{1}\IEEEauthorrefmark{2},
Florian Frank\IEEEauthorrefmark{1}, 
Nicolas Quintin\IEEEauthorrefmark{2},
Fr\'{e}d\'{e}ric Meynadier\IEEEauthorrefmark{1}, \\
Philip Tuckey\IEEEauthorrefmark{1}, 
Anne Amy-Klein\IEEEauthorrefmark{2}
Olivier Lopez\IEEEauthorrefmark{2},
and
Paul-Eric Pottie\IEEEauthorrefmark{1}}
	\IEEEauthorblockA{\IEEEauthorrefmark{1}LNE-SYRTE\\
Observatoire de Paris, Universit\'{e} PSL, CNRS, Sorbonne Universit\'{e},\\
61, avenue de l'Observatoire, 75014 Paris, France\\
Email: paul-eric.pottie@obspm.fr}
\IEEEauthorblockA{\IEEEauthorrefmark{2}Laboratoire de Physique des Lasers,\\
Universit\'{e} Paris 13, CNRS,\\
99 avenue Jean-Baptiste Cl\'{e}ment, Villetaneuse, France}}
\maketitle

\begin{abstract}
We present our work on realizing two-branch fiber links enabling multiple-partner clock comparisons in Europe. We report in detail on the setup connecting two long-haul links, and report for the relative frequency stability and accuracy. We report on a combined uptime of 90$\%$ for our dual-branch link during almost one month. We finally discuss the combined uncertainty contribution of the ensemble of the link architecture to a clock comparison. We show that the frequency transfer uncertainty is $2\times 10^{-19}$.  
\end{abstract}

\section{Introduction}
Comparisons of clocks through fiber links at the continental scale have become a reality in Europe \,\cite{lisdat_clock_2016, guena_first_2017, delva_test_2017}. One challenge is now to combine several coherent fiber links to build a clock network, with the means of comparison making a negligible contribution to the uncertainty compared to the resolution of the clocks. These prospects rely on the ability to realize a multi-branch fiber network. A second challenge is to make the fiber links highly available for on demand optical clocks comparison and for long-period of comparison of fountain cold atoms clocks, over campaign duration of several weeks and months. 

For one aiming at effective optical frequency dissemination service to multi-users, the ability to build optical network and a high signal availability are the key features that are needed. Indeed the comparison arises from the multiplication of the uptime of all the elements constituting the metrology chain. For instance, a two-partite comparison based on 2 clocks, 2 combs and 2 links with uptime of 50$\%$ for each of them leads to a total uptime of only 1.6$\%$. This means that 10 days of operation are needed to obtain 4 hours of non-continuous data. The more partners join the network, the lower the uptime will be, and it makes the use of the dissemination network almost impractical. This calculation illustrates the strong motivation and need to achieve optical link uptime as close as possible to 100$\%$. This requires a high level of automation, remote control, and technical readiness level, but also a delicate control of many experimental aspects when connecting two links together. With that respect, the work presented here is a key step towards a new definition of the SI second based on optical frequency standards and towards novel opportunities for stringent tests of fundamental physics\,\cite{riehle_towards_2015, hees_searching_2016}. 

Another issue that arises with long-duration comparison campaign is the limitation due to the accuracy and the stability of the frequency and time references available for counting the optical beat notes. Fortunately the best performances are not needed everywhere, but at least some level of performances are needed not to loose the benefit of the optical frequency dissemination. Nowadays commercial products based on ultra-stable radio-frequency (RF) oscillator disciplined on Global Positioning System (GPS) are available, and some of them may comply with the needed accuracy in time and frequency. However Global Navigation Satellite System (GNSS) based solutions are not always available. So alternative solutions need to be considered. 

In this paper, we report on the first implementation of a two-branch fiber link, as part of a wider concatenation of links between National Metrological Institutes (NMIs) in Europe enabling multiple clock comparison\,\cite{OFTEN}. We discuss then the results and the limitations due to the RF stability and accuracy.

\section{Experimental set-up}
\begin{figure*}[!t]
  \centering
  \includegraphics[width=.8\textwidth]{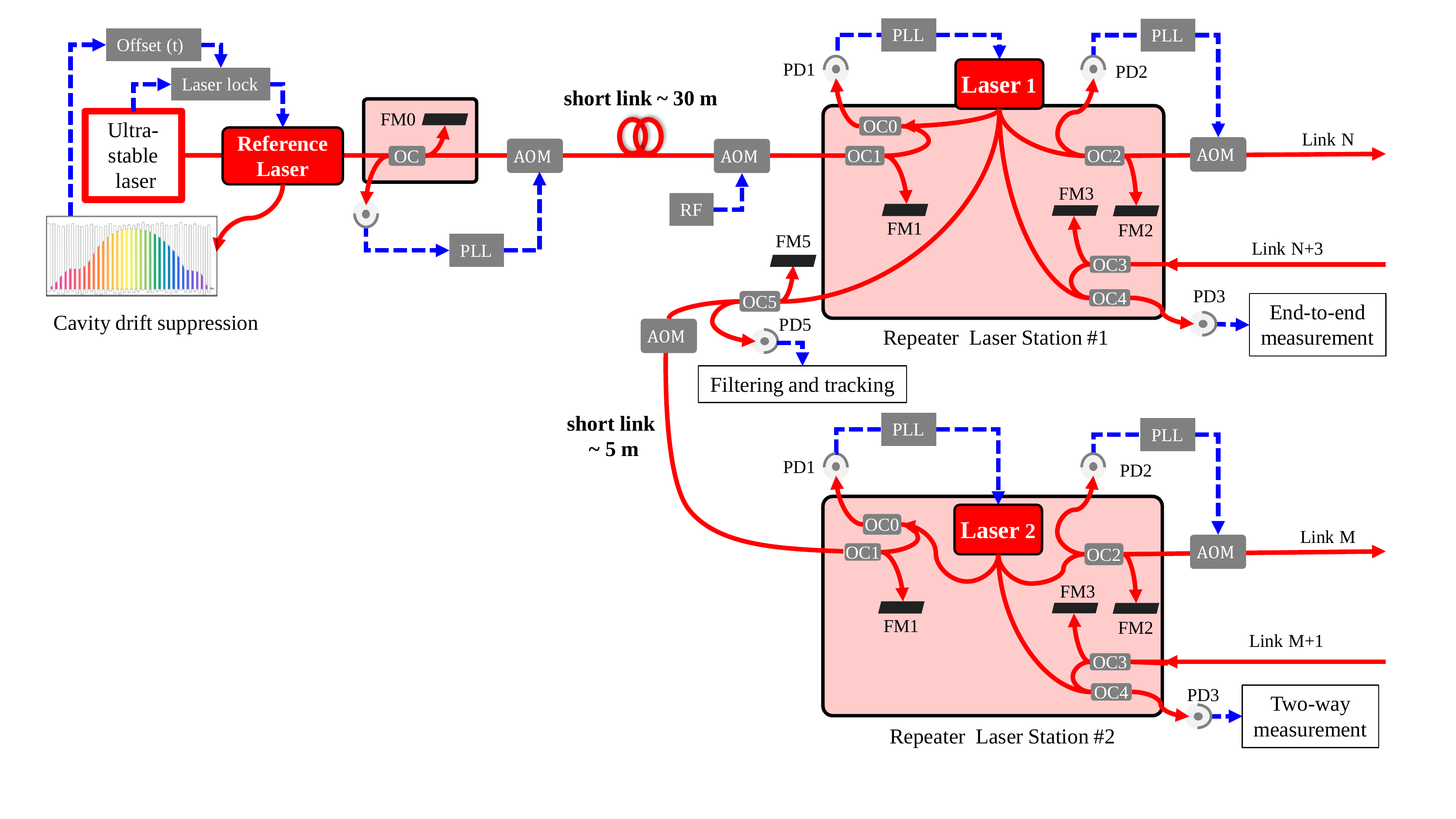}
  \caption{Sketch of the experiment. Upper right link is a four-span link with the first span labelled as $N$ and the 4th span labelled as $N+3$. The lower link on the right is a two-span links with the two spans labelled as $M$ and $M+1$. OC: optical couplers. AOM: Acousto-Optic Modulator. FM: Faraday Miror. PLL: Phase Lock Loop. PD: Photodiode. The light red boxes are thermally controlled boxes. \label{fig:sketch}}  
\end{figure*}

The sketch of the experiment is depicted on Fig.\,\ref{fig:sketch}. We start our experiment with an ultra-stable laser operated at 194.4~THz, continuous wave\,\cite{argence_prototype_2013}. A slave laser is phase-locked to the ultra-stable laser, and actively de-drifted using an optical frequency comb to measure the drift versus an H-Maser \,\cite{letargat_experimental_2013}. The light is then injected into an actively stabilized single-mode fiber of 30-m long and transferred to another experimental room of our laboratory where the links are operated \,\cite{ma_delivering_1994}. Then, we have two repeater laser stations, that are the starting point for two long-haul fiber links\,\cite{chiodo_cascaded_2015, xu_studying_2018}. Each of these links make use of two fibers, one to transfer the optical frequency from our laboratory to the remote point, and the second fiber to evaluate the performance of the frequency transfer. 

The upper link on the right of Fig.\,\ref{fig:sketch} is a 4-span cascaded link of 2$\times$705~km from SYRTE to University of Strasbourg and back\,\cite{lisdat_clock_2016,chiodo_cascaded_2015}. The fiber pair is used for parallel data traffic using the French academic network of RENATER. Therefore the link is equipped with 40 optical add-drop multiplexers (OADMs). 16 bi-directional Erbium-doped fiber amplifiers (EDFAs) compensate partially for the losses, that are about 410~dB, i.e., about 0.29~dB/km. At University of Strasbourg, the signal transferred with this link is compared with the output signal from another long-haul link from PTB in Braunschweig\,\cite{lisdat_clock_2016}. For that purpose, repeater laser stations are operated at the remote end of these two long haul links and the optical frequency comparisons between the international links are done using two-way method between the user-end outputs\,\cite{cantin_progress_2017}, which is useful for the clock comparison campaign\,\cite{lisdat_clock_2016, guena_first_2017, delva_test_2017}.

The lower link on the right of  Fig.\,\ref{fig:sketch} uses a pair of 43-km dedicated fibers from SYRTE to LPL in Villetaneuse. It is a hybrid fiber link, combining active compensation on one fiber and passive compensation on the second fiber, using a two-way beatnote between the local laser and the laser transferred at LPL. Many details of the implementation are reported in\,\cite{xu_studying_2018}. The link is equipped with four OADMs, so that other time and frequency services can be operated in parallel. We are using three bi-directional EDFAs. The losses are about 32~dB in total, i.e., about 0.37~dB/km. At LPL, the transferred frequency signal is compared with the signal transferred from NPL through a 769-km noise-compensated optical link, using two-way frequency comparison\,\cite{delva_test_2017}.

For both upper and lower links on Fig.\,\ref{fig:sketch}, the residual instability is estimated from the end-to-end measurements which are made with the integrated multi-branch Michelson interferometer available in the repeater laser stations\,\cite{chiodo_cascaded_2015,lopez_cascaded_2010}. With this first generation of interferometers built couple of years ago, the length difference between the references arms were not well minimized\,\cite{stefani_tackling_2015}. The length imbalance of the reference arms of the Michelson interferometer of the second repeater laser station is about 15~cm\,\cite{xu_studying_2018}. The remote stations are engineered in such a way that the unaccuracy or instability of their RF reference is not degrading the transferred signal frequency and the end-to-end signal frequency\,\cite{lopez_cascaded_2010}. 

For the specific purpose of a multipartite clock comparison, the frequency of the infra-red (IR) laser sent to the two links has to be precisely chosen so that the beat note frequencies recorded with the remote IR lasers are within the available bandpass filters and the capabilities of the frequency counters. Once the frequencies of the offset phase locks and acousto-optic modulators (AOMs) depicted at the upper part of the sketch Fig.\,\ref{fig:sketch} are chosen, the frequency offset of the lower repeater laser station can be used to tune the beat frequency with the remote IR lasers at the far end of the second link. The issue is to control the optical phase between the two repeater laser stations, that is the frequency fluctuations induced in the short 5-m link between these stations.

We implemented the preliminary solution depicted in Fig.\,\ref{fig:sketch} for compensating these frequency fluctuations. An end-user output of the upper repeater laser station is used to feed the input port of the lower station. The offset frequency of the second repeater laser station is shifted to a well chosen frequency by an AOM. Its RF frequency is 37~MHz. We use the Faraday mirror at the input of the second repeater laser station to reflect back the laser light to the photodiode at the output of the first repeater laser station. On this photodiode, we measure the round-trip delay fluctuations at the carrier  RF frequency of 74~MHz. This signal is tracked and divided by 74, amplified and counted with a dead-time free phase-frequency counter (K-K Messtechnik FXE) with 1-s gate time. It is then subtracted from the clock comparison data performed at the remote end of the link. This solution is a variant of a two-way technique, with one laser light, one AOM and one Faraday mirror. This experimental arrangement is very easy and fast to implement and enables us to connect the two stations without using any servo loops. 

\section{Experimental results}
\begin{figure}[!t]
  \centering
  \includegraphics[width=.45\textwidth]{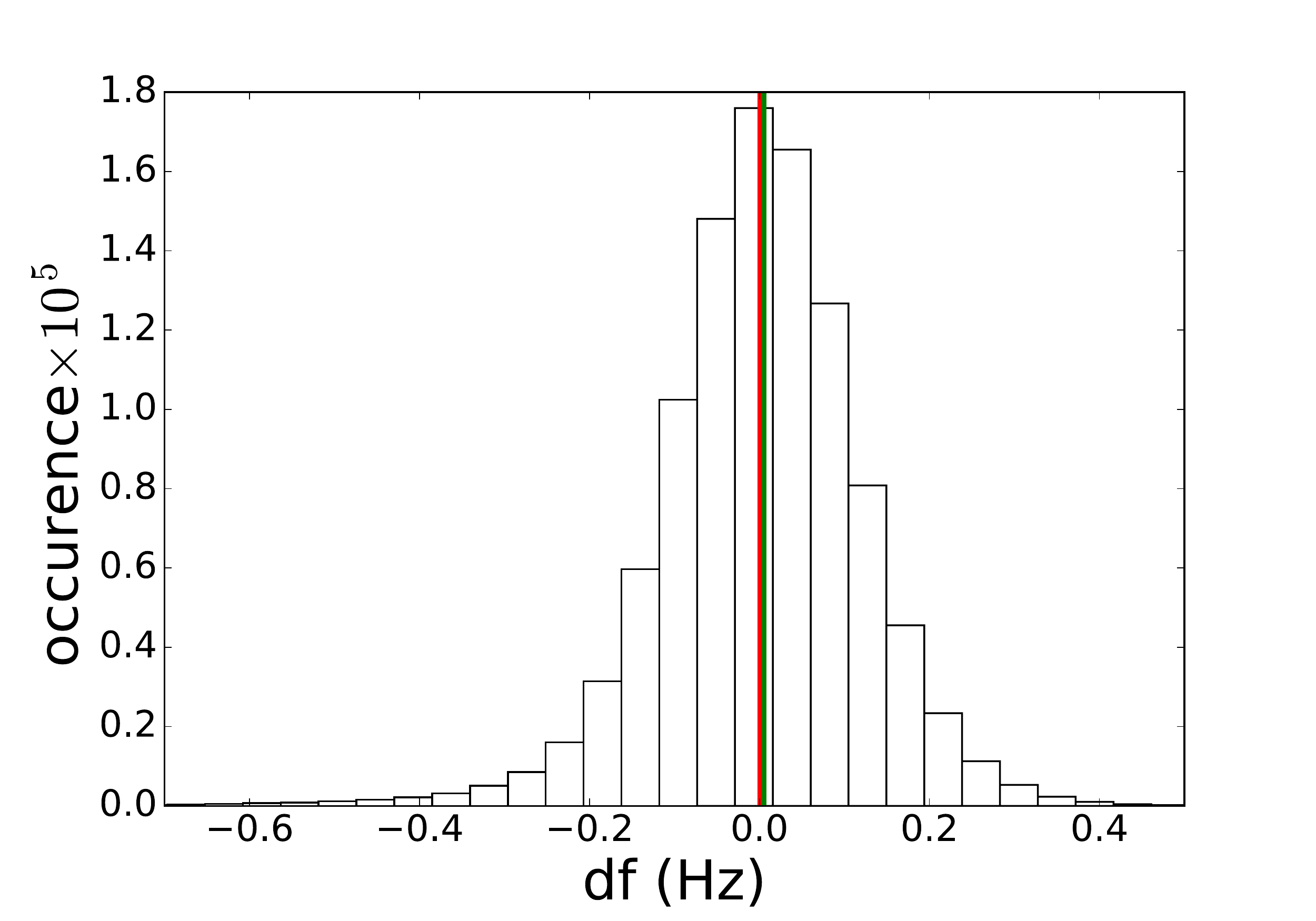}
  \caption{Histogram of the frequency fluctuations $df$ occurring in the 5-m long fiber link, after 12~days of operation, for 1~s averaging time, in Hz. A slight asymmetry is observed. The red line is the mean $m=7.9\times 10^{-19}$. The green light is the median $n=1.7\times 10^{-17}$. Bins width = 45 mHz.\label{Fig:histo}}  
\end{figure}

We now present the data acquired within the frame of the 2017 campaign of comparisons between NPL, SYRTE and PTB, for which we used only one laser source as a seed for two links. We focus on the 5-m short link connecting the two stations. The acquired data represent the free-running round-trip frequency noise of the link to be compensated in post-processing. Figure\,\ref{Fig:histo} shows the histogram for 12 days of continuous data set and a slight asymmetry is observed. One can denote that the tail of the distribution is longer towards smaller frequencies. For a more quantitative analysis, the mean value is $7.9\times 10^{-19}$ and the median is $1.7\times 10^{-17}$. By calculating the rolling mean with a time window of 3600~s, we found that the mean of the relative frequency fluctuation over 3600~s samples varies typically from $\pm 2\times 10^{-17}$ with a daily period, and at worse from $-4\times 10^{-17}$ to $+6\times 10^{-17}$. These fluctuations are two orders of magnitude higher than the residual noise of the long-haul link. These free running fluctuations are fortunately removed in postprocessing. 

\begin{figure}[!t]
  \centering
  \includegraphics[width=.45\textwidth]{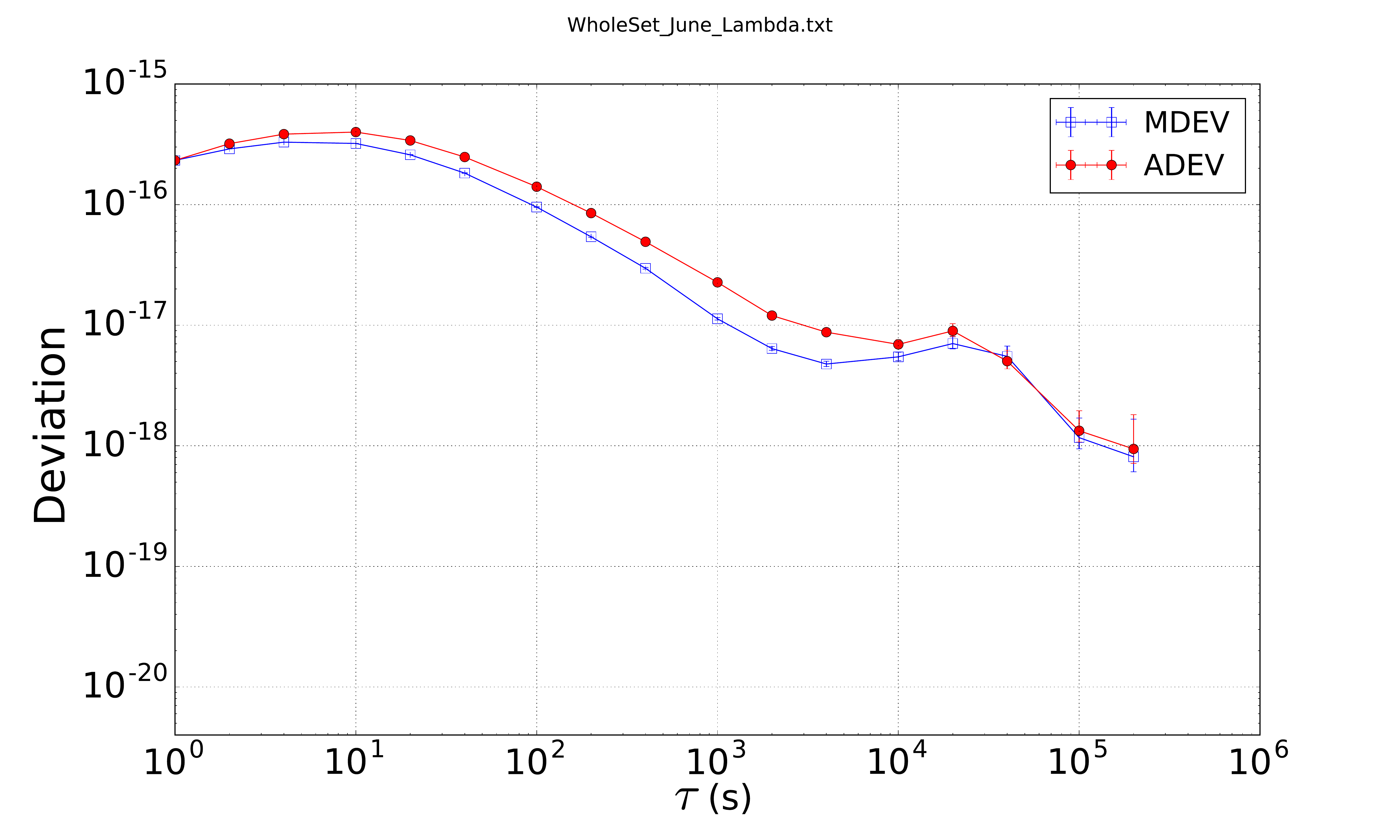}
  \caption{Frequency instability of the 5~m link connecting the two repeater laser stations. Allan standard deviation and modified Allan deviation of the relative frequency fluctuations recorded with 1~s gate time, reporting in $\Lambda$-mode. \label{Fig:devs}}  
\end{figure}

We also present the Allan deviation and Modified Allan deviation of the same 12-days long continuous data sample on Fig.\,\ref{Fig:devs}. We observe that both deviations start at about $2.3\times 10^{-16}$, and reach a maximum of $4\times 10^{-16}$ at 10~s integration time. A clear signature of periodic daily perturbation is to be seen with a local maximum of $8\times 10^{-18}$ at 20\,000~s integration time. The periodic perturbation is due to the temperature of the room that fluctuates daily with a typical amplitude of 1.6~K during this campaign. The signature seen on the Allan deviation is in agreement with the expectation, knowing the temperature fluctuation and the length of the fiber patch-cord. The residual instability after post-processing can be estimated using the link model derived in\,\cite{Newbury_Coherent_2007}, under the assumption that the fiber noise is reciprocal. It results in a post-processed stability of a few $10^{-23}$ at 1~s integration time and around $10^{-30}$ at long-term, which is well below the noise floor of our system. Thus we expect to reach the noise floor of around $10^{-17}$ at 1 s and $10^{-21}$ at long-term. An upper value of this residual noise could also be evaluated from the residual noise of the international frequency comparisons at LPL or University of Strasbourg after post-processing. Uncompensated fiber paths between the Faraday mirror and the coupler in the short 5-m link are also limiting the stability, since this part is only roughly protected against temperature fluctuations. Considering an uncompensated length of 1 m, and estimating residual daily temperature variations of 0.5 K or below, we estimate this limitation to be $4.3\times 10^{-19}$ at half a day averaging time and around $10^{-19}$ at 1 day.

The end-to-end data and their analysis for the two long haul links can be found elsewhere\,\cite{chiodo_cascaded_2015, xu_studying_2018}. For the four-spans link to Strasbourg, the modified Allan deviation is below $10^{-15}$ at 1 s integration time and the long-term stability is typically $10^{-19}$ or below. For the link to LPL, the modified Allan deviation is $6\times 10^{-17}$ at 1 s integration time and the long-term stability is typically $10^{-20}$ at one day integration time. According to\,\cite{chiodo_cascaded_2015, xu_studying_2018}, the frequency bias introduced by the cascaded links and hybrid links are respectively $-4.8\times 10^{-20}$ with an uncertainty of $9\times 10^{-20}$ and $4.2\times 10^{-21}$ with an uncertainty of $8\times 10^{-22}$. The combined uncertainty is therefore conservatively set to $1\times 10^{-19}$. Finally the combined uncertainty of the short and long links and the repeater stations is conservatively set to $2\times 10^{-19}$.

We now discuss the uptime of the multi-branch and composite fiber link, for a duration of 27 continuous days. The various servo loops of the links can indeed unlock or experience some cycle slips due to some short excess noise or during the polarisation optimization in the repeater stations\,\cite{chiodo_cascaded_2015}. The data are summarized in Table\,\ref{Tab:uptime}. The uptime of the 30-m active compensation link connecting the two rooms in our laboratory is 100$\%$. The uptime of the post-processing for the 5-m short link connecting the two stations is assessed by applying a data filtering with a bandwidth of 1~Hz, i.e., about 10 times the standard deviation for integration time ranging from 1~s to 10~s. Since it was installed in a quite noisy environment and not enough protected, it could experience some excess environmental noise which unlocks the tracking used to count the beatnote, leading to an uptime of 98.5$\%$. This shows the need for an operational room with minimum human intervention.

\begin{table}[!t]
\renewcommand{\arraystretch}{1.3}
  \caption{Uptime of all the elements of the two-branch fiber link .  \label{Tab:uptime}}
  \vspace*{2mm}
  \centering
   \begin{tabular}{|l|c|c|c|c|}
   	\hline
	&stabilized link & passive & hybrid &cascaded\\
	\hline
	length (m)		&30 	& 5 & 43\,000 &1\,410\,000\\
	uptime ($\%$) 	&100  &98.5 &95 &96.3\\
	\hline
   \end{tabular}
\end{table}

For the two long-haul links, we assess the uptime from the end-to-end measurements. The hybrid fiber link is assessed from a simple data selection within a 1~Hz bandwidth and removal of outliers for one of the two-way beatnote signals recorded at SYRTE (the TWB3 observable, see details reported in \,\cite{xu_studying_2018}). In the case of our hybrid setup, this is a robust strategy as the short-term noise is about $4\times 10^{-17}$, compared to the hop in frequency of $5\times 10^{-15}$ induced by a single cycle slip. Thus each single cycle slip can be well removed with this procedure and we end up with a conservative estimate of 95$\%$.

\begin{figure}[!t]
  \centering
  \includegraphics[width=.45\textwidth]{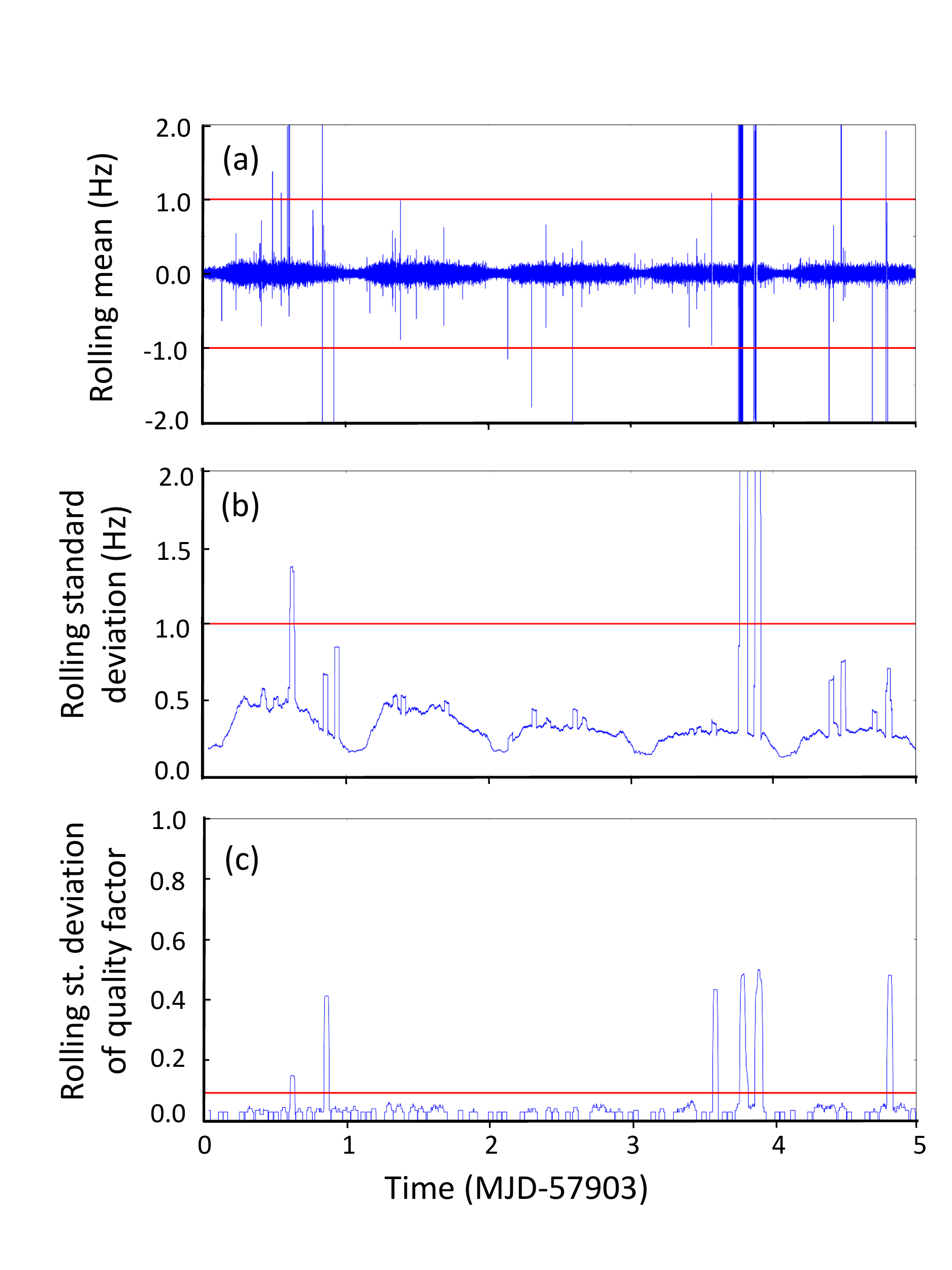}
  \caption{Time evolution of the three observables used for selecting data in the 4-span cascaded link (a) rolling mean of the frequency fluctuations (b) rolling standard deviation of the frequency fluctuations and (c) rolling standard deviation of the quality factor (see text). Red lines indicates the filtering limits. Anomalous noise and outliers are clearly seen.}\label{Fig:rolling}
\end{figure}

Regarding the four-span cascaded link, we developed a more subtle strategy for selecting the data to be kept for a clock comparison. We follow a three-step procedure to select the data. We first apply a very large data filtering with a bandwidth of 10~Hz and produce a boolean quality factor that is set to zero if a point has been removed. Then we process the data and look at the time evolution of three observables, as displayed on Fig.\,\ref{Fig:rolling}. The rolling mean of the relative frequency fluctuation $y$ is performed on a short time of 9~s and allows for detection of outliers. The rolling standard deviation of $y$ is calculated with a longer period of 2750 s in order to detect anomalous noise and reject the period during which the link is not optimally working, and finally the rolling standard deviation of the previously defined quality factor allows for rejection of segment of data where the link is not stable and there is a lot of outliers. This method of data selection was introduced for the purpose of clock comparison and applied on 6 months of data. It seems to be robust as similar performances of the link is found without specific adjustment of the selection procedure parameters. An example of the time evolution of these three observables are displayed on Fig.\,\ref{Fig:rolling} and shows outliers and period of anomalous noise, which are filtered from the data set.

\section{Discussion}
We have shown an easy way to link two repeater laser stations together, with a variant of two-way method, that allows us to match technical requirements of a specific clock comparison campaign. The passive listening of the frequency noise on the short link that connects the two stations reveals asymmetry. These fluctuations are removed from the clock comparison in post processing. It shows that, at the level of exquisite frequency control that optical frequency combs and links are able to achieve, every single detail matters to keep the same level of accuracy. In the specific case of achieving star-liked networks, more complex multi-branches Michelson interferometers were achieved in our groups for the REFIMEVE+ (REseau FIbr\'{e} MEtrologique \`{a} Vocation Europ\'{e}enne+) network, so-called 'super-station', or multi-branches eavesdropping setups as reported in\,\cite{bercy_ultra-stable_2016}. These are clearly better and more robust solutions. However we observe that, if the short link between the two stations would not been monitored, it would be the dominant source of noise and inaccuracy of our set-up. So we believe that this work shines light on the high level of preparation that is needed at the end-user level for a full benefit of the quality of service of optical frequency dissemination network such as REFIMEVE+.  

The combined uptime of the two-branch link is 90$\%$. If the uptime may be compared, this is quite interesting to note that the 1410~km link is now extremely stable and robust, thanks to our efforts for automation and remote monitoring abilities. The uptime of the end-to-end is now probably limited by the uptime of servo loops and set-ups in the clocks and combs in our laboratory that is not fully equipped with automated loops and supervision tools. As compared to previous clock comparisons campaign, for which the uptime of the Paris-Strasbourg link was roughly 90$\%$ but the uptime of the whole Paris-Strasbourg-Pars link was less than 25$\%$, the uptime of our link was considerably improved, and it is comparable to the uptime of atomic fountains as operated at SYRTE and PTB. So now, the strongest limit on the combined uptime of an optical clock comparison is likely to be the uptime of the neutral atom optical clocks themselves.

Further study on the procedure to validate data need to be carried out. A more precise analysis is especially needed to assess carefully the impact on frequency stability and accuracy of the holes made in the raw data set. With the procedure described in this paper, we report a combined accuracy level for the two long-haul links of $1\times 10^{-19}$. This can be compared to the level of accuracy available at remote end points, as in Strasbourg for instance. 

Given that the frequency of the beat note can be up to 55~MHz with the currently used K+K FXE counters, and that the long term stability of commercially available solutions as provided by GPS receiver as the Oscilloquartz OSA 5201 are about $3\times 10^{-13}$ after one day integration time as recorded versus UTC(OP), one obtain a similar contribution to the uncertainty budget of $8.3\times 10^{-20}$. As the performances of the optical clock will improve, that the comparison campaign will last longer, and due to the lack of reliability of some of these commercial GPS receiver, timing solutions provided in-band through the fiber will be needed at some crucial point of the network. White-rabbit solution for instance, considering its excellent level of performance at the level of $1\times 10^{-14}$ at 100~s integration time after 500~km fiber link, and considering its excellent performance-to-cost ratio, seems to us a very promising solution\,\cite{kaur_time_2017}.   

By contrast the requirement on time synchronisation of the measurements over such an optical frequency dissemination network could relax with the time. Basically the inaccuracy imprint on the measurement scales as the ultra-stable laser drift times the de-synchronisation of the measurement, and can be rapidly a dominating source\,\cite{lisdat_clock_2016}. It is quite interesting to notice that the drift of the laser improved by several orders of magnitude over the last five years, from typically a few Hz/s down to sub-mHz/s for the state-of-the-art silicon cavity\,\cite{zhang_ultrastable_2017}. With active de-drift system, 1$\sim$2 mHz/s are achievable nowadays, and it makes the need of accuracy for the synchronisation of the measurement up to the 50~ms level for the same level of accuracy of $1\times 10^{-19}$. 

\section{Conclusion}
We have shown a two-branch long-haul fiber link that enables the comparison of multiple optical clocks at their best level of performances to date. We have shown that the two-branch fiber link is not limiting the uptime of the clock comparison. This work represents a decisive step towards the realization of a robust fiber network for clock comparisons, contributing to progress along the roadmap towards the redefinition of the SI second. It also impacts the future use of the REFIMEVE+ network, since the benefit of ultra-stable signal also need good experimental control at the end-user side. We have also presented discussions on the requirements for RF reference signal and time reference signal, and show that the most stringent need are for the RF accuracy. Solutions based on White-Rabbit and/or other in-band techniques, far from being a burden, may open new opportunities for time and RF standard dissemination.

\section*{Acknowledgment} We thank Rodolphe Le Targat and Hector Alvarez-Martinez for the operation of the ultra-stable laser and the comb at start of our experiment. We thank Michel Abgrall and Baptiste Chupin for allowing us to measure GPS receiver versus UTC(OP). We would like to thank E. Camisard, T. Bono and L. Gyd\'{e} for their support and for facilitating the access to the RENATER network and facilities, P. Gris and B. Moya of Uni. Strasbourg. We acknowledge funding support from the Agence Nationale de la Recherche (Labex First-TF ANR-10-LABX-48-01, Equipex REFIMEVE ANR-11-EQPX-0039, Idex PSL ANR-10-IDEX-0001-02), the European Metrology Programme for Innovation and Research (EMPIR) under SIB-02 OFTEN. EMPIR is jointly funded by the EMPIR participating countries within EURAMET and the European Union.

\end{document}